\documentclass[aps,prb,reprint,preprintnumbers,superscriptaddress,amsmath,amssymb,bibnotes,longbibliography]{revtex4-2}
\usepackage{graphicx}
\usepackage{dcolumn}
\usepackage{epstopdf}
\usepackage{bm,multirow,threeparttable,soul}
\usepackage{flushend}
\usepackage[pdfstartview=FitH, CJKbookmarks=true, bookmarksnumbered=true, bookmarksopen=true, colorlinks=true, pdfborder=true, linkcolor=blue, anchorcolor=blue, citecolor=blue,urlcolor=blue,breaklinks=true]{hyperref}
\usepackage[mathlines]{lineno}

\begin{document}
\title{Nodeless superconductivity in noncentrosymmetric LaRhSn}
\author{Z. Y. Nie}
\affiliation{Center for Correlated Matter and School of Physics, Zhejiang University, Hangzhou 310058, China}
\affiliation  {Zhejiang Province Key Laboratory of Quantum Technology and Device, Department of Physics, Zhejiang University, Hangzhou 310058, China}
\author{J. W. Shu}
\affiliation{Center for Correlated Matter and School of Physics, Zhejiang University, Hangzhou 310058, China}
\affiliation  {Zhejiang Province Key Laboratory of Quantum Technology and Device, Department of Physics, Zhejiang University, Hangzhou 310058, China}
\author{A. Wang}
\affiliation{Center for Correlated Matter and School of Physics, Zhejiang University, Hangzhou 310058, China}
\affiliation  {Zhejiang Province Key Laboratory of Quantum Technology and Device, Department of Physics, Zhejiang University, Hangzhou 310058, China}
\author{H. Su}
\affiliation{Center for Correlated Matter and School of Physics, Zhejiang University, Hangzhou 310058, China}
\affiliation  {Zhejiang Province Key Laboratory of Quantum Technology and Device, Department of Physics, Zhejiang University, Hangzhou 310058, China}
\author{W. Y. Duan}
\affiliation{Center for Correlated Matter and School of Physics, Zhejiang University, Hangzhou 310058, China}
\affiliation  {Zhejiang Province Key Laboratory of Quantum Technology and Device, Department of Physics, Zhejiang University, Hangzhou 310058, China}
\author{A. D. Hillier}
\affiliation{ISIS Facility, STFC Rutherford Appleton Laboratory, Harwell Science and Innovation Campus, Oxfordshire, OX11 0QX, United Kingdom}
\author{D. T. Adroja}
\affiliation{ISIS Facility, STFC Rutherford Appleton Laboratory, Harwell Science and Innovation Campus, Oxfordshire, OX11 0QX, United Kingdom}
\affiliation{Highly Correlated Matter Research Group, Physics Department, University of Johannesburg, P.O. Box 524, Auckland Park 2006, South Africa}
\author{P. K. Biswas}
\affiliation{ISIS Facility, STFC Rutherford Appleton Laboratory, Harwell Science and Innovation Campus, Oxfordshire, OX11 0QX, United Kingdom}
\author{T. Takabatake}
\affiliation{Center for Correlated Matter and School of Physics, Zhejiang University, Hangzhou 310058, China}
\affiliation{Department of Quantum Matter, AdSE, Hiroshima University, Higashi-Hiroshima 739-8530, Japan}
\author{M. Smidman}
\email[Corresponding author: ]{msmidman@zju.edu.cn}
\affiliation{Center for Correlated Matter and School of Physics, Zhejiang University, Hangzhou 310058, China}
\affiliation  {Zhejiang Province Key Laboratory of Quantum Technology and Device, Department of Physics, Zhejiang University, Hangzhou 310058, China}
\author{H. Q. Yuan}
\email[Corresponding author: ]{hqyuan@zju.edu.cn}
\affiliation  {Center for Correlated Matter and School of Physics, Zhejiang University, Hangzhou 310058, China}
\affiliation  {Zhejiang Province Key Laboratory of Quantum Technology and Device, Department of Physics, Zhejiang University, Hangzhou 310058, China}
\affiliation  {State Key Laboratory of Silicon Materials, Zhejiang University, Hangzhou 310058, China}

\date{\today}

\begin{abstract}
The superconducting order parameter of the noncentrosymmetric superconductor LaRhSn is investigated by means of low temperature measurements of the specific heat, muon-spin relaxation/rotation ($\mu$SR)  and the tunnel-diode oscillator (TDO) based method. The specific heat and magnetic penetration depth [$\lambda(T)$] show an exponentially activated temperature dependence, demonstrating fully gapped superconductivity in LaRhSn. The temperature dependence of $\lambda^{-2}(T)$ deduced from the TDO based method and $\mu$SR show nearly identical behavior, which can be well described by a single-gap \emph{s}-wave model, with a zero temperature gap value of $\Delta(0)=1.77(4)k_BT_c$. The zero-field $\mu$SR spectra do not show detectable changes upon cooling below $T_c$, and therefore there is no evidence for time-reversal-symmetry breaking in the superconducting state.

\begin{description}
\item[PACS number(s)]

\end{description}
\end{abstract}

\maketitle
\section{INTRODUCTION}
Noncentrosymmetric superconductors (NCS) have attracted considerable interest, since in the absence of inversion symmetry, an antisymmetric potential gradient gives rise to an antisymmetric spin-orbit coupling (ASOC). The ASOC lifts the  two-fold spin degeneracy of the electronic bands,  potentially allowing for unconventional superconducting properties such as the admixture of spin-singlet and spin-triplet pairing states \cite{mixture,ASOC}.  In the noncentrosymmetric heavy fermion superconductor CePt$_3$Si, measurements of the magnetic penetration depth, thermal conductivity and specific heat showed the presence of line nodes in the energy gap \cite{CePt3Silinenode,CePt3Si2,CePt3Si3}, and nodal superconductivity was subsequently found in other NCS, such as Li$_2$Pt$_3$B \cite{Li2Pt3B,Li2Pt3B1NMR}, Y$_2$C$_3$ \cite{Y2C32}, K$_2$Cr$_3$As$_3$ \cite{K2Cr3As3Cao,K2Cr3As3Pang}, and ThCoC$_2$ \cite{Bhatt2019}. However, many NCS are found to be fully gapped superconductors, such as Mo$_3$Al$_2$C \cite{Bonalde2011,Bauer2014}, $RT$Si$_3$ ($R$ = La, Sr, Ba, Ca; $T$ = transition metal) \cite{Bauer2009,LaTSi31,LaTSi32,Kneidinger2014,Eguchi2011}, BiPd \cite{BiPdJL2,BiPd2NC}, Re$_6T$ \cite{Re6ZrRPSingh,Re6Zr1pang,Re6Hf}, La$_7T_3$ \cite{La7Ir3,La7Rh3}, BeAu \cite{Amon2018} and PbTaSe$_2$ \cite{PbTaSe21,PbTaSe22,Wilson2017}. Even though some of these systems have been found to have multiple superconducting gaps, many NCS show evidence for single gap \emph{s}-wave superconductivity, indicating negligible contributions from a spin-triplet pairing component. The predominance of such $s$-wave superconductivity even in systems with strong ASOC has posed the question as to what conditions are required to give rise to  mixed parity pairing. In addition, even in NCS exhibiting unconventional properties, unambiguosly demonstrating the presence of singlet-triplet mixing remains challenging, and obtaining direct evidence  may require probing associated topological superconducting phenomena such as gapless edge modes and Majorana modes \cite{Sato2009,Sato2010}.

Time reversal symmetry breaking (TRSB) has been observed in the superconducting states of some weakly correlated NCS, such as LaNiC$_2$ \cite{LaNiC2TRS}, La$_7T_3$ \cite{La7Ir3,La7Rh3}, and several Re-based superconductors \cite{Re6ZrRPSingh,Re24Ti5,Re24Nb5}. TRSB has primarily been revealed by muon-spin relaxation measurements, which detect the spontaneous appearance of small magnetic fields in the superconducting state, even in the absence of external applied fields \cite{Ghosh2020}. In most cases, such systems have been found to have nodeless superconducting gaps, which has often been difficult to reconcile with the unconventional nature of the pairing state implied by TRSB. On the other hand, different behavior was recently found in the weakly correlated NCS CaPtAs, where there is evidence for both nodal superconductivity and TRSB \cite{CaPtAs1,CaPtAs2}. Consequently, it is important to survey a wide range of different classes of NCS, so as to look for novel behaviors arising from ASOC, as well as to reveal the origin of any time reversal symmetry breaking and to understand its relationship to the broken inversion symmetry.

\begin{figure}
\includegraphics[angle=0,width=0.49\textwidth]{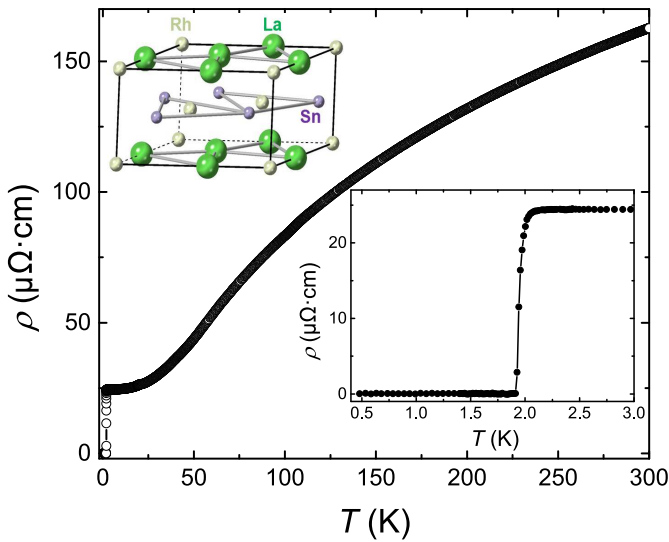}
\vspace{-12pt} \caption{\label{figure1}(Color online) Temperature dependence of the electrical resistivity $\rho(T)$ of LaRhSn from room temperature down to 0.5~K. The insets show $\rho(T)$ near the superconducting transition, and the crystal structure of LaRhSn.}
\vspace{-12pt}
\end{figure}

LaRhSn crystallizes in the noncentrosymmetric hexagonal ZrNiAl-type structure (space group \emph{P$\bar{6}2$m}) displayed in the inset of Fig.~\ref{figure1}, where the rare-earth atoms form a distorted kagome lattice. Compounds in this family with a magnetic rare-earth atom have been extensively studied due to the interplay of strong electronic correlations and frustrated magnetism \cite{Zhao2019,Tokiwa2015,CeIrSn}, while several other systems with nonmagnetic rare-earth elements are superconductors. For example, Sc(Ir,Rh)P, LaRhSn, LaPdIn are superconductors with relatively low transition temperatures $T_{\rm c}$ \cite{LaRhSn1,ScIrP,ScRhP,LaPdIn}, while (Zr,Hf)RuP, ZrRu(As,Si) and Mo(Ni,Ru)P have $T_c$'s over 10~K \cite{ZrHfRuP,ZrRuAs,ZrRuAs2,ZrRuSi,MoNiRuP}, where the higher $T_c$ values may be a consequence of the phonon spectra and electron-phonon coupling strengths  \cite{ZrRuAs2,HfIrSi,ZrRuPDFT}. In this article, we study the order parameter of LaRhSn via  measurements of the electronic specific heat and magnetic penetration depth, where the latter is probed  using both the tunnel-diode oscillator (TDO) based method and muon-spin rotation ($\mu$SR). The experimental results obtained by various techniques can be consistently described by a single-gap \emph{s}-wave model corresponding to weak electron-phonon coupling. In addition, zero-field $\mu$SR measurements do not exhibit detectable changes below $T_c$, and therefore there is no evidence for TRSB in the superconducting state.

\section{EXPERIMENTAL DETAILS}

Single crystals of LaRhSn were synthesized using the Czochralski method, as described in Ref \onlinecite{LaRhSn3}. The specific heat was measured in a Quantum Design Physical Property Measurement System (PPMS) with a $^3$He insert. The resistivity $\rho(T)$ was measured in a $^3$He cryostat from room temperature down to 0.5~K, using a standard four-probe method. $\mu$SR measurements were performed using the MuSR spectrometer at the ISIS pulsed muon source of the Rutherford Appleton Laboratory, UK \cite{MuSR,MuSRdata}. The $\mu$SR experiments were conducted in transverse-field (TF) and zero-field (ZF) configurations, so as to probe the flux line lattice (FLL) and the presence or absence of time-reversal symmetry breaking, respectively. Powdered single crystals of LaRhSn were mounted on a high-purity silver sample holder, which was mounted on a dilution refrigerator, with a temperature range from 0.05~K to 2.5~K. With an active compensation system, the stray magnetic field at the sample position can be canceled to within 1 $\mu$T.  TF-$\mu$SR experiments were carried out in several fields up to 60~mT.

The shift of the magnetic penetration depth from the zero-temperature value $\Delta\lambda(T)=\lambda(T)-\lambda(0)$ was measured down to 0.3~K in a $^3$He cryostat, using a tunnel-diode oscillator (TDO) based method \cite{TDOdevice,Gfactor,TDO2}, with an operating frequency of 7~MHz and a noise level of 0.1~Hz. Samples with typical dimensions of $550 \times 450 \times 300$ $\mu$m$^3$, were mounted on a sapphire rod. The generated ac field is about 2 $\mu$T, which is much smaller than the lower critical field $H_{c1}$, ensuring that the sample remains in the Meissner state.  $\Delta\lambda(T)$ is proportional to the frequency shift from zero temperature $\Delta f(T)$, i.e., $\Delta\lambda(T)$ = $G\Delta f(T)$, where $G$ is the calibration factor determined from the geometry of the coil and  sample \cite{Gfactor}.

\section{RESULTS}
\subsection{Electrical resistivity and specific heat}

\begin{figure}
\includegraphics[angle=0,width=0.49\textwidth]{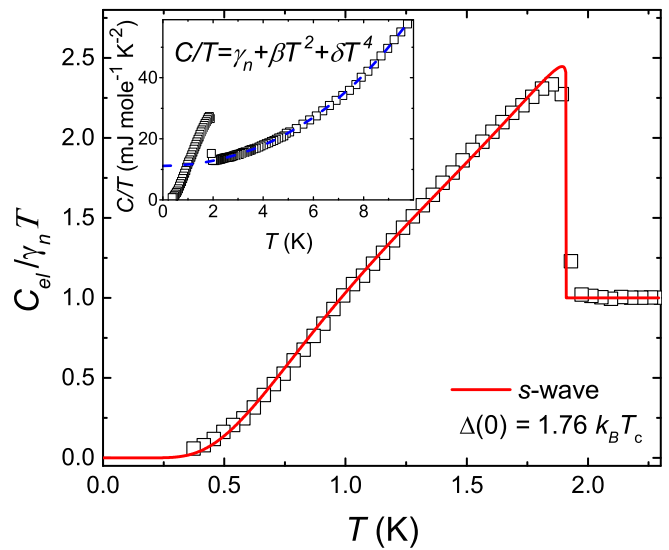}
\vspace{-12pt} \caption{\label{figure2}(Color online) Temperature dependence of the electronic specific heat as $C_{el}(T)/\gamma_n$\emph{T} of LaRhSn, where the solid line represents fitting with a single-gap \emph{s}-wave model. The inset displays the total specific heat $C(T)/T$, where the dashed line represents the fitting to the normal state contribution.}
\vspace{-12pt}
\end{figure}

The single crystals of LaRhSn were characterized by measurements of the electrical resistivity and specific heat. Figure~\ref{figure1} displays the electrical resistivity $\rho(T)$ from room temperature down to 0.5~K, which exhibits metallic behavior in the normal state. The inset shows  $\rho(T)$ at low temperatures, where there is a sharp superconducting transition at around 2.0~K.

\begin{figure}
\includegraphics[angle=0,width=0.49\textwidth]{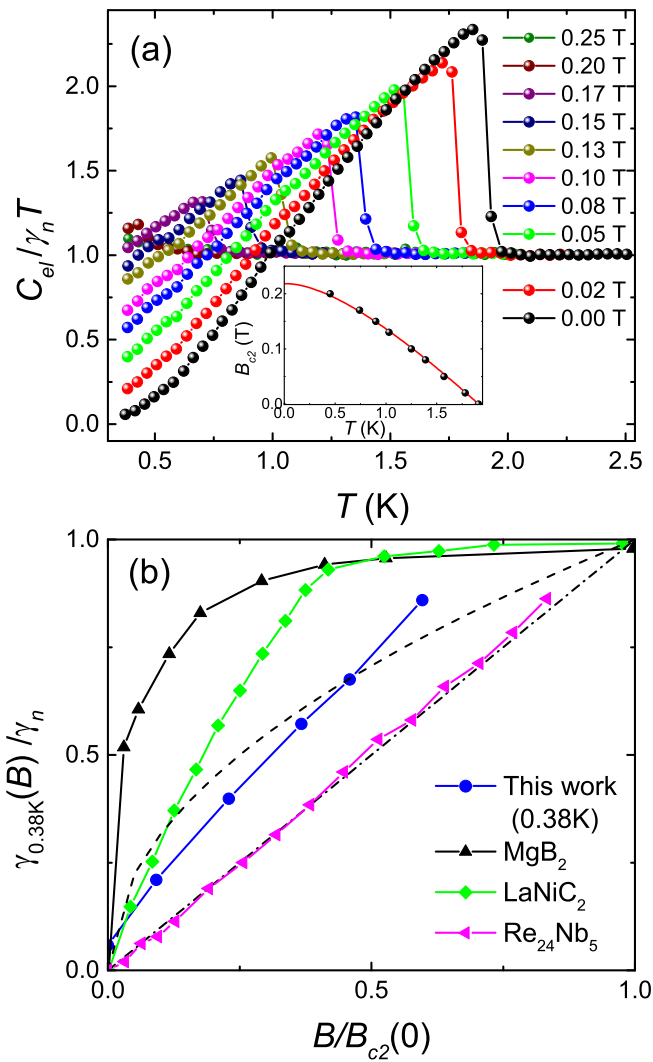}
\vspace{-12pt} \caption{\label{figure3}(Color online) (a) Temperature dependence of the electronic specific heat as $C_{el}/\gamma_n$\emph{T} of LaRhSn under various applied fields. The inset displays the temperature dependence of the upper critical field $B_{c2}(T)$, derived from the specific heat measurements, where the solid line represents fitting with the WHH model where $B_{c2}(0)= 0.219(2)$~T. (b) Field dependence of the residual Sommerfeld coefficient plotted as $\gamma_{0.38\mathrm{K}}(B)$/$\gamma_n$ versus $B/B_{c2}(0)$ for LaRhSn, Re$_{24}$Nb$_{5}$ \cite{Re24Nb5}, MgB$_2$ \cite{MgB2} and LaNiC$_2$ \cite{LaNiC22}. The dashed and  dashed-dotted lines correspond to the expected behaviors of nodal and single-gap \emph{s}-wave superconductivity, respectively.}
\vspace{-12pt}
\end{figure}

The inset of Figure. \ref{figure2} displays the total specific heat $C(T)/T$ of LaRhSn in zero field, where there is a clear superconducting transition with a midpoint $T_c$~=~1.9~K, in line with the behavior of $\rho(T)$. In the normal state, the specific heat data are fitted by $C(T)/T = \gamma_n +\beta T^2 + \delta T^4$, with $\gamma_n$~=~11.15(4)~mJ mole$^{-1}$~K$^{-2}$, $\beta$~=~0.410(6)~mJ mole$^{-1}$~K$^{-4}$ and $\delta$~=~0.87(1)~$\mu$J mole$^{-1}$ K$^{-6}$. Here $\gamma_n$ is the normal state Sommerfeld coefficient, and the latter two terms represent the phonon contribution. The Debye temperature $\theta_D$ is estimated to be 241(1)~K using $\theta_D$~=~$(12\pi^4Rn/5\beta)^{1/3}$, where $R$~=~8.31 J mole$^{-1}$ K$^{-1}$ is the molar gas constant and \emph{n}~=~3 is the number of atoms per formula unit. The electron-phonon coupling constant $\lambda_{\textrm{el-ph}}$ can be approximated via
\begin{equation}
\lambda_{\textrm{el-ph}}~=~\frac{1.04+\mu^*\textrm{ln}(\frac{\theta_D}{1.45T_c})}{(1-0.62\mu^*)\textrm{ln}(\frac{\theta_D}{1.45T_c})-1.04}.
\label{equation1}
\end{equation}
Using the typical values for $\mu^*$ of 0.1 -- 0.15, $\lambda_{\textrm{el-ph}}$~=~0.47 -- 0.57 are obtained, close to the derived values for isostructural LaPdIn \cite{LaPdIn}, indicating weakly coupled superconductivity in LaRhSn. In addition, the value of $\gamma_n$ is very similar to that of LaPdIn, but larger than the values for LuPdIn and LaPtIn which are not superconducting down to at least 0.5~K \cite{LaPdIn}. This is consistent with the magnitude of the density of states at the Fermi level playing an important role in giving rise to superconductivity in this family of compounds.

The main panel of Fig.~\ref{figure2} shows the low temperature electronic specific heat $C_{el}(T)/\gamma_n T$, from which  the phonon contribution has been subtracted. In the superconducting state, the entropy \emph{S} can be calculated by \cite{MgB22}

 \begin{equation}
  S~=~-\frac{3\gamma_n}{\pi^3}\int_0^{2\pi}\int_0^\infty[f\textrm{ln}f+(1-f)\textrm{ln}(1-f)]d\varepsilon  d\phi,
 \label{equation2}
 \end{equation}
where the $f(E,T)$~=~[1+exp(\emph{E}/$k_BT)]^{-1}$ is the Fermi-Dirac distribution function. Here, $E=\sqrt{\varepsilon^2+\Delta_k^2}$, where $\Delta_k(T)=\Delta(T)g_k$ is the superconducting gap function. Therefore, the electronic specific heat of superconducting state can be obtained by $C_{el}=TdS/dT$.  In the case of a single-gap \emph{s}-wave model, there is no angle dependent component ($g_k=1$), and $\Delta(T)$ was approximated by \cite{delta0}
\begin{equation}
\Delta(T)~=~\Delta(0){\rm tanh}\left\{1.82\left[1.018\left(T_c/T-1\right)\right]^{0.51}\right\},
\label{equation3}
\end{equation}
\noindent where $\Delta(0)$ is the zero-temperature superconducting gap magnitude. As shown by the solid line in  Fig. \ref{figure2}, the zero field $C_{el}/\gamma_n T$ can be well described by this single-gap \emph{s}-wave model, with $\Delta(0)=1.76(1)k_BT_c$.

Upon applying a magnetic field, the bulk superconducting transition is shifted to lower temperatures and is completely suppressed at about 0.25~T (see Fig. \ref{figure3} (a)). The inset displays the extracted upper critical field $B_{c2}(T)$  and the corresponding fitting using the Werthamer-Helfand-Hohenberg (WHH) model \cite{WHH}, with a zero temperature upper critical field $B_{c2}(0)=0.219(2)$~T. Using $\lambda(0)=\sqrt{\Phi_0B_{c2}(0)}/\sqrt{24\gamma_n}\Delta(0)$ \cite{lamda0}, where the units of $B_{c2}(0)$ and $\gamma_n$ are gauss and ergs~cm$^{-3}$~K$^{-2}$, respectively, a penetration depth at zero temperature $\lambda(0)=244(1)$~nm is estimated using $\Delta(0)=1.76(1)k_BT_c$. Combined with a Ginzburg-Landau (GL) coherence length of $\xi_{GL}=\sqrt{\Phi/2\pi B_{c2}(0)}=38.7(2)$~nm, the GL parameter $\kappa$ is estimated to be 6.30(4), indicating that LaRhSn is  a type-\uppercase\expandafter{\romannumeral2} superconductor. Using the values of $\lambda(0)$=244(1)~nm, a residual normal state resistivity $\rho_0= 25~\mu\Omega$~cm and $\gamma_n=11.15(4)$ mJ mole$^{-1}$K$^{-2}$,  the mean free path $\ell$ and BCS coherence length $\xi_{\mathrm{BCS}}$ are estimated to be $\ell$=17.91(8)~nm and $\xi_{\mathrm{BCS}}$=43.8(2)~nm \cite{V3Si}. The mean free path $\ell$ is smaller than $\xi_{\mathrm{BCS}}$, indicating that the sample is in the dirty limit.

Figure \ref{figure3} (b) displays the field dependence of the Sommerfeld coefficient value at 0.38~K, normalized by its value in the normal-state, i.e., $\gamma_{0.38\mathrm{K}}(B)$/$\gamma_n$. It can be seen that $\gamma_{0.38\mathrm{K}}(B)$/$\gamma_n$ shows a nearly linear field dependence, being similar to the fully gapped superconductor Re$_{24}$Nb$_{5}$ \cite{Re24Nb5}. On the other hand, $\gamma_{0.38\mathrm{K}}(B)$/$\gamma_n$ clearly deviates from the square-root field dependence (dashed line)  expected for line nodal superconductors, as well as the typical behaviors of the multiband superconductors MgB$_2$ \cite{MgB2} and LaNiC$_2$ \cite{LaNiC22}. Note that $\gamma_{0.38\mathrm{K}}(B)$/$\gamma_n$ of LaRhSn are determined from  the specific heat at the lowest measured temperature, and therefore even in zero-field the data have a finite value.

\begin{figure}
\includegraphics[angle=0,width=0.49\textwidth]{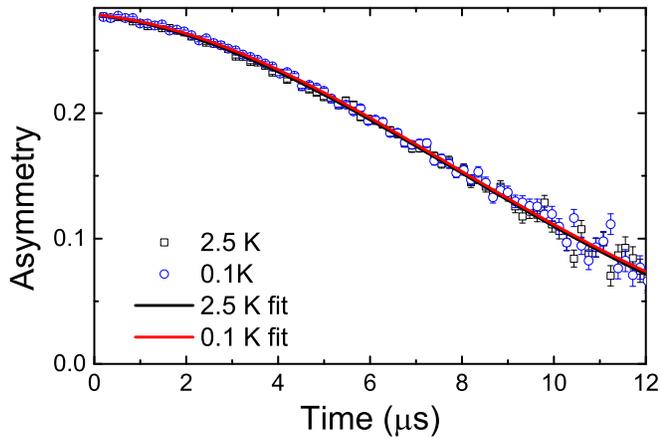}
\vspace{-12pt} \caption{\label{figure4}(Color online) ZF-$\mu$SR spectra of LaRhSn at 2.5~K ($T>T_c$) and 0.1~K ($T<T_c$). The solid lines show the results from  fitting using Eq. \ref{equation4}.}
\vspace{-12pt}
\end{figure}
\subsection{$\mu$SR measurements}

Figure \ref{figure4} displays the zero-field (ZF) $\mu$SR spectra collected at 2.5~K ($T>T_c$) and 0.1~K ($T<T_c$). These are fitted with a damped Gaussian Kubo-Toyabe (KT) function
\begin{equation}
G_\textrm{ZF}(t)=A\left[\frac{1}{3}+\frac{2}{3}(1-\delta^2t^2)\textrm{exp}\left(-\frac{\delta^2t^2}{2}\right)\right]\textrm{exp}(-\Lambda t)+A_{\textrm{bg}},
\label{equation4}
\end{equation}
\noindent where \emph{A} is the initial asymmetry, and $A_{\textrm{bg}}$ corresponds to the time independent background term from muons stopping in the silver sample holder. $\delta$ and $\Lambda$ are the Gaussian and Lorentzian relaxation rates, respectively. Upon fitting with Eq. \ref{equation4}, $\delta=0.086(3)~\mu{\rm s}^{-1}$ and $\Lambda=0.0134(11)~\mu{\rm s}^{-1}$ were obtained at 2.5 K, while $\delta=0.082(3)~\mu{\rm s}^{-1}$ and $\Lambda=0.0157(10)~\mu{\rm s}^{-1}$ at 0.1~K. Therefore, we find no evidence for TRSB in the superconducting state of LaRhSn, and these results suggest that any spontaneous internal fields should be no larger than 6.6 $\mu$T, which is smaller than the corresponding fields in other reported TRSB superconductors \cite{Ghosh2020}.

\begin{figure}
\includegraphics[angle=0,width=0.49\textwidth]{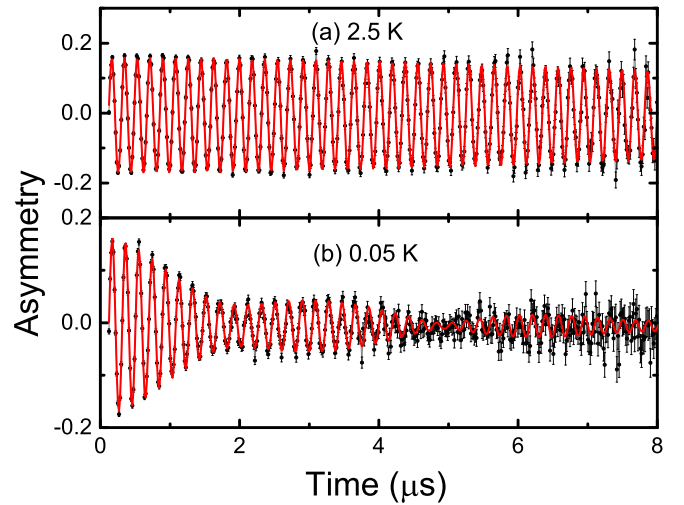}
\vspace{-12pt} \caption{\label{figure5}(Color online) Transverse field $\mu$SR spectra of LaRhSn at (a) 2.5~K ($T>T_c$) and (b) 0.05~K ($T<T_c$) in an applied field of 40~mT. The solid lines show the results of fitting with Eq.~\ref{equation5}}
\vspace{-12pt}
\end{figure}

\begin{figure}
\includegraphics[angle=0,width=0.49\textwidth]{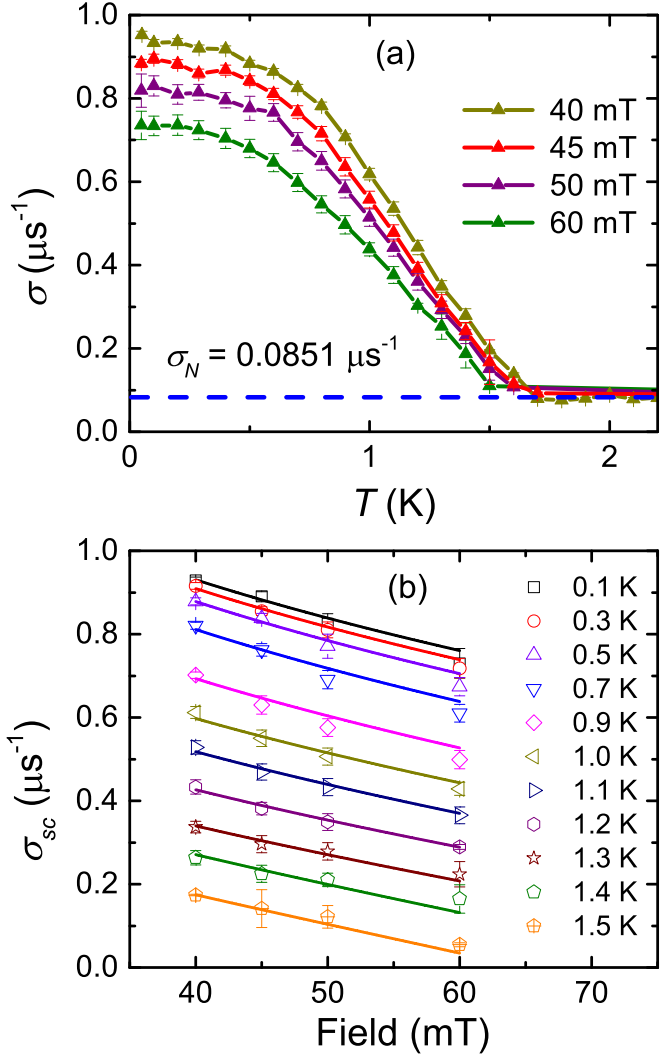}
\vspace{-12pt} \caption{\label{figure6}(Color online) (a) Temperature dependence of the Gaussian relaxation rate of the TF-$\mu$SR spectra in different applied fields between 40~mT and 60~mT. (b) Field dependence of the superconducting contribution to the TF-$\mu$SR relaxation rate $\sigma_{sc}$ at various temperatures, where the solid lines correspond to fitting using Eq. \ref{equation8}.}
\vspace{-12pt}
\end{figure}

Transverse-field $\mu$SR (TF-$\mu$SR) measurements were carried out in the mixed state with applied fields in the range 40~mT to 60~mT, where the data were collected upon field-cooling in order to probe a well-ordered flux-line lattice (FLL). The results at 2.5~K and 0.05~K in a field of 40~mT are displayed in Fig.~\ref{figure5}. The significant increase of the depolarization rate corresponds to the inhomogeneous field distribution in the sample, characteristic of the formation of a FLL. The TF-$\mu$SR asymmetry were fitted to the sum of oscillations damped by Gaussian decaying functions
 \begin{equation}
  G_\textrm{TF}(t)~=~\sum_{i=1}^nA_i\textrm{cos}(\gamma_{\mu}B_it+\phi)e^{-(\sigma_it)^2/2} + A_{\textrm{BG}},
\label{equation5}
\end{equation}

\noindent where $A_i$ is the amplitude  of the oscillating component, which precesses about a local field $B_i$ with a common phase offset $\phi$ and a Gaussian decay rate $\sigma_i$, while $\gamma_{\mu}/2\pi$~=~135.5 MHz/T and $A_{\textrm{BG}}$ are the muon gyromagnetic ratio and background term, respectively. The asymmetry  can be well fitted with three oscillatory components ($n$~=~3), where $\sigma_3$ was fixed to zero, corresponding to muons stopping  in the silver sample holder. Figure \ref{figure6}(a) displays the temperature dependence of $\sigma(T)$ obtained following the multiple-Gaussian method  described in Ref \onlinecite{multigaussian}. Here, the first and second moment of the field distribution are calculated as
\begin{equation}
  \langle B\rangle~=~\sum_{i=1}^{n-1}\frac{A_i~B_i}{A_1+\cdot\cdot\cdot A_{n-1}},
\label{equation6}
\end{equation}

\begin{equation}
  \langle B^2\rangle~=~\sum_{i=1}^{n-1}\frac{A_i}{A_1+\cdot\cdot\cdot A_{n-1}}[(\sigma_i/\gamma_\mu)^2+[B_i - \langle B\rangle]^2],
\label{equation7}
\end{equation}
\noindent and $\sigma=\gamma_{\mu}\sqrt{\langle B^2\rangle}$. The relaxation rate in the normal state is ascribed to a temperature independent contribution arising from quasistatic  nuclear moments, with a nuclear dipolar relaxation rate $\sigma_N~=~0.0851(27)~\mu s^{-1}$. The superconducting component of the variance $\sigma_{sc}$ is calculated as $\sigma_{sc}$~=~$\sqrt{\sigma^2-\sigma_N^2}$, and its  field dependence is displayed in Fig. \ref{figure6}(b) for several temperatures.

 For  small applied fields and large $\kappa$, $\sigma_{sc}$ is field independent and proportional to $\lambda^{-2}$, which is not applicable for the current measurements of LaRhSn. On the other hand, for $\kappa \geq5$ and 0.25/$\kappa^{1.3}$ $\leq$ $b$ $\leq$ 1, $\sigma_{sc}$ may  be approximated by~\cite{sigmafit}
\begin{equation}
  \sigma_{sc}~=~4.854 \times 10^4\frac{1}{\lambda^2}(1-b)[1+1.21(1-\sqrt{b})^3],
\label{equation8}
\end{equation}
where \emph{b}~=~$B/B_{c2}$ is the applied field normalized by the upper critical field. Since the $\kappa$ of LaRhSn was determined to be about 6.30(4), the measurements of LaRhSn are within the applicability of Eq. \ref{equation8}. Therefore by fixing $B_{c2}(T)$ to the bulk values derived from the specific heat in Fig.~\ref{figure3}, the temperature dependence of $\lambda^{-2}(T)$ can be obtained from fitting with Eq.~\ref{equation8} [Fig.~\ref{figure6}(b)], and the results are shown in Fig. \ref{figure8}, together  with the TDO results described in following section.

\begin{figure}
\includegraphics[angle=0,width=0.49\textwidth]{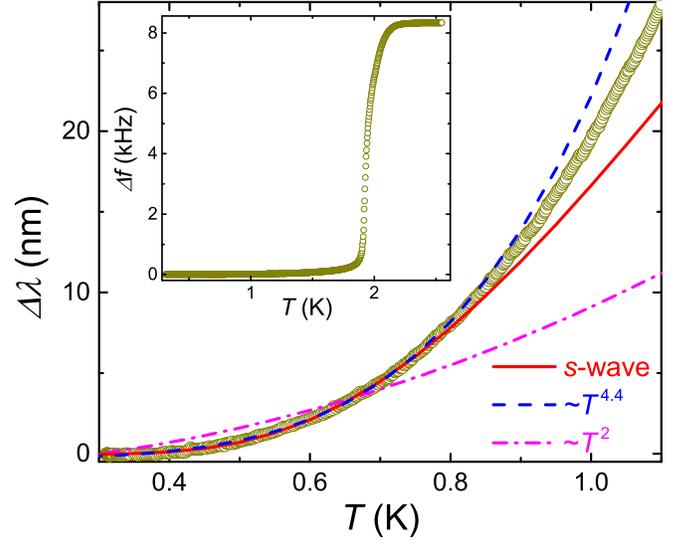}
\vspace{-12pt} \caption{\label{figure7}(Color online) The change of magnetic penetration depth $\Delta\lambda(T)$ of LaRhSn at low temperatures. The solid red, dashed blue and dashed-dotted magenta lines represent fitting to an \emph{s}-wave model, and power-law dependences $\sim T^{4.4}$ and $\sim T^2$, respectively. The inset displays the frequency shift $\Delta f(T)$ from 2.5~K down to 0.3~K, where there is  a sharp superconducting transition at around $T_c$~=~2~K.}
\vspace{-12pt}
\end{figure}

\subsection{TDO measurements and superfluid density analysis}

Figure \ref{figure7} shows the penetration depth shift $\Delta\lambda(T)$ of LaRhSn at low temperatures, with a calibration factor \emph{G}~=~14.2~$\textrm{{\AA}/Hz}$. The inset displays the frequency shift $\Delta f(T)$ from 2.5~K down to the base temperature of 0.3~K, where a sharp superconducting transition is observed at $T_c=2$~K, in accordance with other measurements. Upon further cooling, $\Delta\lambda(T)$ flattens at the lowest measured temperatures, indicating fully gapped superconductivity in LaRhSn. For an \emph{s}-wave superconductor, the temperature dependence of $\Delta\lambda(T)$ for $T\ll T_c$ can be approximated by

 \begin{equation}
\Delta\lambda(T)=\lambda(0)\sqrt{\frac{\pi\Delta(0)}{2k_BT}}\textrm{exp}\left(-\frac{\Delta(0)}{k_BT}\right).
\label{equation9}
\end{equation}

\noindent As shown by the solid line, the experimental data below $T_c/3$ can be well described by the \emph{s}-wave model with $\Delta(0)=1.80(1)k_BT_c$, where $\lambda(0)=227.9$~nm was fixed to the value derived from TF-$\mu$SR. The data were also fitted by a power law dependence $\Delta\lambda(T)\propto T^n$, from 0.3~K up to 0.75~K. A large exponent of \emph{n}~=~4.4 is obtained, which is much larger than two, excluding nodal superconductivity in LaRhSn.

\begin{figure}
\includegraphics[angle=0,width=0.49\textwidth]{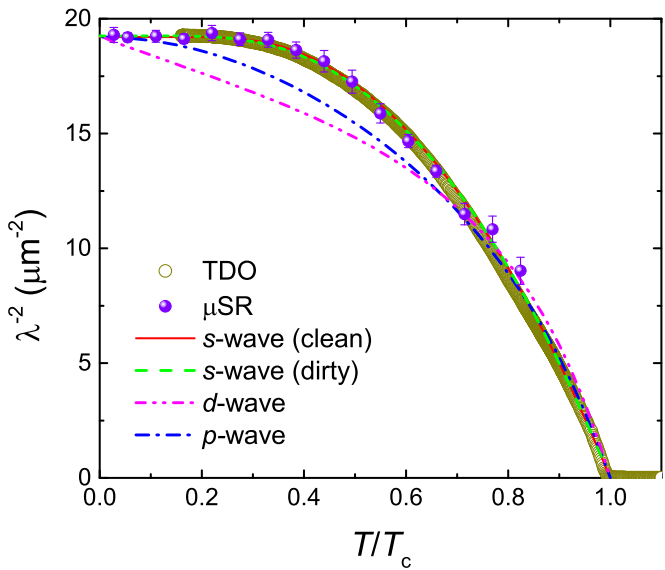}
\vspace{-12pt} \caption{\label{figure8}(Color online) Temperature dependence of $\lambda^{-2}(T)$ as a function of the normalized temperature $T/T_c$. The data are derived from measurements using the TDO based method and TF-$\mu$SR measurements,  which correspond to the empty circle and solid symbols, respectively. The lines show the results from fitting with different models for the gap structure.}
\vspace{-12pt}
\end{figure}

To further characterize the superconducting pairing state of LaRhSn,  the temperature dependence of $\lambda^{-2}(T)$ was analyzed, which is proportional to the superfluid density $\rho_s(T)$ as $\rho_s(T)=[\lambda(0)/\lambda(T)]^2$. Figure \ref{figure8} displays $\lambda^{-2}(T)$ as a function of the reduced temperature $T/T_c$, where the data are derived from both the TDO and TF-$\mu$SR measurements, which show nearly identical behavior. Since the previous analysis suggested that the sample is in the dirty limit, the results from TF-$\mu$SR were fitted with the following expression for a dirty \emph{s}-wave model \cite{dirtys}

\begin{equation}
\rho_s(T)~=~\frac{\Delta(T)}{\Delta(0)}\textrm{tanh}\left\{\frac{\Delta(T)}{2k_BT}\right\}.
\label{equation11}
\end{equation}

\noindent As shown by the dashed line in Fig. \ref{figure8}, the dirty \emph{s}-wave model can well describe the experimental data, with $\lambda(0)=227.9(9)~$nm and $\Delta(0)=1.77(4)k_BT_c$. The data were also analyzed using the clean limit expression
 \begin{equation}
\rho_{\rm s}(T)=\frac{\lambda^{-2}(T)}{\lambda^{-2}(0)}=1+2 \left\langle\int_{\Delta_k}^{\infty}\frac{EdE}{\sqrt{E^2-\Delta_k^2}}\frac{\partial f}{\partial E}\right\rangle_{\rm FS},
\label{equation10}
\end{equation}\noindent where a clean single gap \emph{s}-wave model can also fit the data well, yielding a larger gap value of $\Delta(0)=2.05(3)k_BT_c$. Here the gap value obtained from the dirty \emph{s}-wave model is in very good agreement to those derived from the analysis of specific heat and low temperature $\Delta\lambda(T)$, while the clean limit value is considerably larger, which is in-line with the previous dirty limit calculation. We note that due to the samples being in the dirty limit, we cannot exclude an anisotropic superconducting gap in LaRhSn, since impurity scattering can suppress any gap anisotropy. On the other hand, as also shown in Fig. \ref{figure8}, a \emph{d}-wave model with $g_k$~=~cos~2$\phi$ and \emph{p}-wave model with $g_k$~=~sin~$\theta$ ($\phi$= azimuthal angle, $\theta$= polar angle) cannot account for the data, further indicating a lack of nodal superconductivity in LaRhSn. Meanwhile, the value of $\lambda(0)$ obtained from $\mu$SR experiments is very close to that from  specific heat results. Using this value of $\lambda(0)=227.9(9)$nm, $\kappa=5.89(4)$ is estimated, which corresponds well to the value from the specific heat analysis. The obtained superconducting parameters of LaRhSn are  displayed in Table~\ref{ResTab}. Therefore, the results of specific heat, TDO-based measurements and $\mu$SR  can all be consistently described by a single-gap \emph{s}-wave model with a gap magnitude very close to that of weak-coupling BCS theory, and there is no evidence for time-reversal symmetry breaking below $T_{\rm c}$.

\begin{table}[t]
\centering
\caption{Superconducting parameters of LaRhSn, where the parentheses with $C$ and $\mu$SR denote the results from the specific heat and  $\mu$SR, respectively.}
\label{ResTab}
\begin{ruledtabular}
\begin{threeparttable}
\resizebox{\columnwidth}{!}{
\begin{tabular}{c c c}
\multirow{1}*{Property}&\multirow{1}*{Unit}&\multicolumn{1}{c}{Value}\\[2pt]
\cline{1-3}
\\[1pt]
$T_{c}$&K&1.9\\[2pt]
$B_{c2}$(0)&T&0.219(2)\\[2pt]
$\gamma_n$&mJ mole$^{-1}$K$^{-2}$&11.15(4)\\[2pt]
$\Theta_D$&K&241(1)\\[2pt]
$\lambda_{el-ph}$& &0.47-0.57\\[2pt]
$\xi_{GL}$&nm&38.7(2)\\[2pt]
$\ell$&nm&17.91(8)\\[2pt]
$\xi_{\mathrm{BCS}}$&nm&43.8(2)\\[2pt]
$\lambda_{0}$(C)&nm&244(1)\\[2pt]
$\lambda_{0}$$(\mu\mathrm{SR})^{\mathrm{dirty}}$&nm&227.9(9)\\[2pt]
$\kappa$(C)& &6.30(4)\\[2pt]
$\kappa$$(\mu\mathrm{SR})^{\mathrm{dirty}}$& &5.89(4)\\[2pt]
$\Delta(0)$(C)&$k_BT_c$&1.76(1)\\[2pt]
$\Delta(0) $$(\mu\mathrm{SR})^{\mathrm{dirty}}$&$k_BT_c$&1.77(4)\\[2pt]
\end{tabular}
}

     \end{threeparttable}
\end{ruledtabular}
\end{table}

\section{SUMMARY}
In summary, we have studied the order parameter of the noncentrosymmetric superconductor LaRhSn. Both the specific heat and magnetic penetration depth show exponentially activated behavior at low temperatures, providing strong evidence for fully gapped superconductivity. $\lambda^{-2}(T)$ derived from the TDO based method and TF-$\mu$SR, as well as the specific heat can be consistently well described by a single-gap \emph{s}-wave model, with a gap magnitude very close to that of weak coupling BCS theory. Together with findings for LaPdIn \cite{LaPdIn} and ZrRuAs \cite{ZrRuAs2}, our results suggest that fully gapped \emph{s}-wave superconductivity, together with a lack of evidence for time reversal symmetry breaking, are consistent common features of weakly correlated NCS with the ZrNiAl-type structure and there is a lack of significant singlet-triplet mixing.

\begin{acknowledgments}
This work was supported by the National Key R$\&$D Program of China (Grant No.~2017YFA0303100), the Key R$\&$D Program of Zhejiang Province, China (Grant No.~2021C01002), the National Natural Science Foundation of China (Grant No.~11874320, No.~11974306 and No.~12034017), and the Zhejiang Provincial Natural Science Foundation of China (R22A0410240). D.T.A. would like to thank the Royal Society of London for Advanced Newton Fellowship founding between UK and China. Experiments at the ISIS Pulsed Neutron and Muon Source were supported by a beamtime allocation from the Science and Technology Facilities Council (Grant No.~RB2010190 \cite{MuSRdata})

\end{acknowledgments}

%
\end{document}